\documentclass[oupdraft]{bio}
\usepackage[colorlinks=true, urlcolor=citecolor, linkcolor=citecolor, citecolor=citecolor]{hyperref}

\usepackage{amsmath}
\usepackage{graphicx,psfrag,epsf}
\usepackage{enumerate}
\usepackage{natbib}
\usepackage{url} 
\usepackage{amssymb,amsmath,color,bm}
\usepackage{graphics,subfigure}
\usepackage[figuresright]{rotating}
\usepackage{bbold}

\newcommand{\y}{\mathbf{y}}
\newcommand{\Y}{\mathbf{Y}}
\newcommand{\x}{\mathbf{x}}
\newcommand{\X}{\mathbf{X}}

\newcommand{\bc}{\mathbf{b}}
\newcommand{\B}{\mathbf{B}}
\newcommand{\T}{\intercal}
\newcommand{\w}{\mathbf{w}}
\newcommand{\vv}{\mathbf{v}}
\newcommand{\UU}{\mathbf{U}}
\newcommand{\V}{\mathbf{V}}

\begin{document}

\title{\vspace{-40 pt} Discriminating sample groups with multi-way data}

\author{TIANMENG LYU\\
	\textit{Division of Biostatistics, School of Public Health,
		University of Minnesota,}\\
		 \textit{Minneapolis, MN 55455, USA}\\[4pt]
	ERIC F. LOCK$^\ast$\\
	\textit{Division of Biostatistics, School of Public Health,
		University of Minnesota,}\\
		 \textit{Minneapolis, MN 55455, USA}\\
		{elock@umn.edu}\\[4pt]
	LYNN E. EBERLY\\
	\textit{Division of Biostatistics, School of Public Health,
		University of Minnesota,}\\
		 \textit{Minneapolis, MN 55455, USA}\\
}

\markboth%
{T. Lyu and others}
{Discriminating sample groups with multi-way data}

\maketitle
\footnotetext{To whom correspondence should be addressed.}

\begin{abstract}
{High-dimensional linear classifiers, such as the support vector machine (SVM) and distance weighted discrimination (DWD), are commonly used in biomedical research to distinguish groups of subjects based on a large number of features.  However, their use is limited to applications where a single vector of features is measured for each subject.  In practice data are often \emph{multi-way}, or measured over multiple dimensions.  For example, metabolite abundance may be measured over multiple regions or tissues, or gene expression may be measured over multiple time points, for the same subjects.    We propose a framework for linear classification of high-dimensional multi-way data, in which coefficients can be factorized into weights that are specific to each dimension.  More generally, the coefficients for each measurement in a multi-way dataset are assumed to have low-rank structure.  This framework extends existing classification techniques, and we have implemented multi-way versions of SVM and DWD. We describe informative simulation results, and apply multi-way DWD to data for two very different clinical research studies.  The first study uses metabolite magnetic resonance spectroscopy data over multiple brain regions to compare patients with and without spinocerebellar ataxia, the second uses publicly available gene expression time-course data to compare treatment responses for patients with multiple sclerosis.  Our method improves performance and simplifies interpretation over naive applications of full rank linear classification to multi-way data.  An R package is available at \url{https://github.com/lockEF/MultiwayClassification}.}
{Classification; Distance weighted discrimination; Gene time-course; magnetic resonance spectroscopy; Support vector machine; Tensors}
\end{abstract}

\section{Introduction}
\label{sec:intro}

In biomedical research and other fields, data are often best represented as a \emph{multi-way} array, also called a \emph{tensor}.  A multi-way array simply extends the familiar two-way data matrix (e.g., \emph{Samples} $\times$ \emph{Variables}) to higher dimensions.  Multi-way data frequently arise from molecular profiling and imaging modalities, where data may be measured over multiple body regions, tissue-types, or developmental time points.  Our motivating example is magnetic resonance spectroscopy (MRS) measurement of the abundance of several metabolites in three brain regions for a common set of participants: \emph{samples} $\times$ \emph{metabolites} $\times$ \emph{regions}.  We also consider gene expression time-course data, in which the expression of many genes are measured over multiple time points: \emph{samples} $\times$ \emph{genes} $\times$ \emph{times}.

There are a large number of exploratory factorization and dimension reduction techniques for multi-way data.  A detailed survey of these methods can be found in \cite{kolda2009tensor}. Two classical methods are the PARAFAC \citep{harshman1970foundations} decomposition and the Tucker \citep{tucker1966some} decomposition, which extend well-known methods such as the singular value decomposition and principal component analysis for a data matrix.  These and similar factorization techniques are frequently used in practice to analyze neuroimaging data \citep{cichocki2013} and in other biostatistical applications \citep{allen2012, zhou2015}.

In addition to exploratory approaches, there is also a small but growing literature on supervised methods for multi-way data, where the interest is to determine the relationship between an outcome vector and covariates that have multi-way structure. \citet{zhou2013tensor} proposed tensor regression models which have a continuous clinical outcome as the outcome variable and images that are covariates, formulated as multi-way arrays.  In their model, covariate coefficients are assumed to have a PARAFAC structure. An analogous Bayesian formulation for tensor regression models is described by \citet{miranda2015}.

Our interest is in classification of a categorical outcome from high-dimensional multi-way data. Classification methods that identify a hyperplane that provides linear separation between two classes are commonly used in biomedical research to distinguish groups of subjects based on several features, but these methods assume that each sample's predictors are in a vector; the methods can thus not be applied to multi-way data where each sample's predictors are a matrix.  There has been some work to extend classifiers to the multi-way context in machine learning and computer vision.  \cite{ye2004two} extended the traditional Fisher's linear discriminant analysis (LDA) to two-way tensors and their application focused on dimension reduction of images. \cite{bauckhage2007robust} extended the least mean squares approach for LDA to tensors by assuming that the projection tensor can be given by the PARAFAC model. \cite{tao2005supervised} proposed a supervised tensor learning scheme which can be applied using different learning methods such as support vector machines (SVM) (\cite{cortes1995support}) and LDA; in their formulation, coefficients for the hyperplane can be factorized into a single set of weights that are specific to each dimension (a rank-$1$ factorization).

Much of the previous work on supervised learning from multi-way data focuses on the tensor version of LDA. However, LDA can result in overfitting and also the solution is not identifiable when the number of predictors is larger than the sample size, which is frequently the case for high-throughput biomedical data. An alternative is SVM,  which identifies a high-dimensional hyperplane that separates two classes. The hyperplane is chosen to maximize distance between cases and controls that are closest to the hyperplane; these samples define the support vectors. But as shown in \cite{marron2007distance}, SVM may suffer from the data piling problem, which means if we project the data onto the normal vector of the separating hyperplane then many points will pile up at the margin. In order to overcome the data piling problem in SVM, \cite{marron2007distance} proposed the Distance Weighted Discrimination (DWD) method, which allows every data point to affect the estimation of the hyperplane.

In this article, we describe a general framework for classifying high-dimensional multi-way data that extends existing linear classification approaches.  Our central assumption is that the multi-way coefficient matrix can be decomposed into patterns that are particular to each dimension, giving a low-rank representation. The coefficients are estimated by iteratively updating the weights in each dimension to optimize an objective function.  This is shown to improve both interpretation and performance over naive applications of linear classification that ``vectorize" each sample's multi-way structure and treat each array entry as a separate variable.  We implement our extended versions of both SVM and DWD for multi-way data, and find that DWD generally performs better.  In applications we illustrate how cross-validation can be used for model assessment, and how bootstrapping can be used to assess the uncertainty of model estimates.

Previous work on supervised tensor learning has been primarily motivated by imaging data, and our method can be applied to images.  However, we are primarily motivated by applications where each dimension of a multi-way array has a unique interpretation, and a rank-1 or low-rank model makes intuitive sense.  For the \emph{Metabolites} $\times$ \emph{Regions} data introduced above, we apply multi-way DWD to distinguish patients with spinocerebellar ataxia type I (SCA1) from controls; for the \emph{genes} $\times$ \emph{times} data we apply multi-way DWD to distinguish good and poor responders to IFN$\beta$ treatment for multiple sclerosis \citep{baranzini2005}. In both cases multi-way DWD improves performance over the naive approach and allows for a simpler interpretation of the results.

\section{Methods}
\label{sec:meth}

\subsection{High-dimensional classification}

Here we briefly describe linear classification based on a high-dimensional covariate vector per sample, before discussing the multi-way case in Section~\ref{MultiWayFull}.

Suppose data are available for $n$ subjects, each belonging to one of two classes which we denote by $+1$ and $-1$. Let $\x_{i}$ denote the $p\times 1$ vector of covariates for subject $i$, and let $y_{i}$ denote the class labels  $y_i\in\{+1,-1\}$, $i=1,\ldots,n$. Define $\y = [y_{1}, \ldots, y_n]: 1 \times n$ and $\X = [\x_{1}, \hdots, \x_n]: p \times n$. The goal is to find the hyperplane $\bc=[b_1,\ldots,b_p]^{\T}:p \times 1$ which best separates the two classes. That is, the projections $f(\x_i)=\x_i^{\T} \bc$ should provide good discrimination between the two classes.  Performance is assessed via an objective function $h(\y,\X,\bc,\Theta)$, which is to be minimized.  The exact form of the objective function $h$ and additional parameters $\Theta$ (if any) depend on the method.  Below we briefly describe the objective functions for SVM and DWD, respectively.

\emph{SVM objective:}  SVM uses the \emph{hinge loss}
function.  The optimization problem can be formulated as
\begin{equation*}
\underset{\bc,\beta}{\mbox{argmin}} \frac{1}{n} \left[ \sum_{i} \max (0,1-y_i(\x_i^{\T} \bc-\beta)) \right] + \lambda||\bc||^2,
\end{equation*}
where $\beta$ is an intercept term and $\lambda$ is a penalty parameter that determines the tradeoff between the size of the hyperplane margin and correct classification of the groups on either side of the hyperplane.

\emph{DWD objective:} In contrast to SVM, DWD allows every data point (sample) to influence $\bc$ by optimizing the sum of the inverse distances from the data points to the hyperplane. Let $\Y$ be the $n\times n$ diagonal matrix with $y_{i}$'s as the diagonal components.  The optimization problem in DWD can be formulated as
\begin{equation*}
\underset{\textbf{\textit{r}},\bc,\beta,\bm{\xi}}{\mbox{argmin}}\sum_{i}\frac{1}{r_i}+C\textbf{1}'\bm{\xi},
\end{equation*}
where $\textbf{1}$ is the vector of $1$'s, $C$ is the penalty parameter and $\bm{\xi}$ is a penalized vector with the following constraints:
\begin{equation*}
\textbf{\textit{r}}=YX'\bc+\beta \textbf{y}+\bm{\xi} \ge 0, \, \, \, \, \parallel \bc\parallel\le 1, \, \, \bm{\xi}\ge 0.
\end{equation*}

\subsection{Naive (full) multi-way classification}

\label{MultiWayFull}
Now consider classification of samples with array data $samples \times dim_1 \times dim_2$. Let $x_{ijk}$ denote the value of measurement under the $j$th characteristic of $dim_1$ and the $k$th characteristic in $dim_2$ for subject $i$ where $i = 1, \ldots, n$, $j = 1, \ldots, p$ and $k = 1, \ldots, m$. A naive approach to extend linear classifiers to multi-way data is to estimate the coefficient $b_{jk}$ for each $dim_{1,j} \times dim_{2,k}$ pair in the data array. Let \[\x_{i}=[x_{i11}, \ldots, x_{ip1}, x_{i12},\ldots, x_{ip2}, \ldots, x_{i1m}, \ldots, x_{ipm}]^\T: pm \times 1\] denote the full vector of covariates for subject $i$, $i=1,\ldots,n$. Then the hyperplane that is used to distinguish the two classes is
\begin{equation*}
f(\x_{i})=b_{11}x_{i11}+b_{12}x_{i12}+\cdots+b_{1m}x_{i1m}+\cdots+b_{p1}x_{ip1}+b_{p2}x_{ip2}+\cdots+b_{pm}x_{ipm},
\end{equation*}
We define this naive approach as the full model and let $\B: p \times m$ be the coefficient array with $b_{jk}$ as the $(j,k)$th component.

\subsection{Rank 1 multi-way classification}
\label{MultiWay1}
The full model in Section~\ref{MultiWayFull} estimates different coefficients for the same $j$th characteristic in $dim_1$ under different characteristics in $dim_2$. For example, if $dim_1$ corresponds to \emph{metabolites} and $dim_2$ corresponds to brain \emph{regions}, then the full model estimates different coefficients for the same metabolite, e.g., glucose, measured in different brain regions. But the effects of the same metabolite in different brain regions on the classification of the two classes are very likely correlated. The full model does not account for the known multi-way structure of the data and ignores the possible correlation among the different $dim_{1,j} \times dim_{2,k}$ pairs; hence, it may result in less accurate classification performance. The proposed multi-way model can be regarded as a low rank approximation of the full model. The rank 1 multi-way model has the simplest form among all of the low rank approximations and has a very straightforward interpretation: The model assumes that the coefficient matrix $\B_{p\times m}$ has the rank $1$ decomposition
\begin{equation*}
\B_{p\times m}=\w \vv^\T,
\end{equation*}
where $\w=[w_{1}, \ldots, w_{p}]^\T$ denotes the vector of weights for $dim_1$ and $\vv=[v_{1}, \ldots, v_{m}]^\T$ denotes the vector of weights for $dim_2$. Under this assumption, the hyperplane to separate the two classes is:
\begin{equation}
\begin{array}{rcl}
f(\X_{i})&=&w_{1}v_{1}x_{i11}+w_{1}v_{2}x_{i12}+\ldots+w_{p}v_{1}x_{ip1}+w_{p}v_{2}x_{ip2}+\ldots+w_{p}v_{m}x_{ipm} \\
&=&(\vv^\T  \X_i^\T)\w\\
&=&(\w^\T \X_i)\vv , 
\end{array}\label{MultiWayPlane}
\end{equation}
where $w_{j}, j=1,\ldots,p$ represents the weight on the $j$th characteristic of $dim_1$ and $v_{k}, k=1,\ldots,m$ represents the weight on the $k$th characteristic of $dim_2$. Since we estimate the weights specific to each dimension, and a larger absolute weight usually implies a more important characteristic in terms of its influence on classification, we interpret the importance of different characteristics in one dimension to be proportional across each level of the other dimension. The full model does not assume any commonality to the effects of characteristics in $dim_1$ across the levels of $dim_2$, since the coefficients are estimated separately.

\subsection{Rank $r$ multi-way classification}
\label{MultiWayR}
The rank 1 multi-way model assumes that the coefficient matrix has a rank 1 decomposition, but sometimes the rank 1 structure may not be able to represent all the information in the true coefficient matrix. For example, in the $metabolites \times regions$ example described in Section~\ref{sec:intro}, the rank 1 model assumes that there is only one distinguishing profile of metabolites ($\vv$) but it can be weighted differently across the different regions ($\w$). However, in reality, the truth might be that there are multiple distinguishing metabolite profiles ($\vv_1,\vv_2,\cdots,\vv_r$),  which should be weighed differently across the different regions ($\w_1,\w_2,\cdots,\w_r$). Under such circumstances, we need a more complicated model compared to the rank 1 multi-way model. We propose a rank $r$ model for the coefficient matrix $\B_{p\times m}$, which can be viewed as a compromise between the full model and the rank 1 multi-way model. The rank $r$ multi-way model assumes that the coefficient matrix has the following decomposition:
\begin{equation}
\B_{p\times m}=\w_{1} \vv_{1}^\intercal+ \cdots +\w_{r} \vv_{r}^\intercal, \label{rankReq}
\end{equation}
where $\w_{z}=[w_{z1}, \ldots, w_{zp}]^\intercal$ and $\vv_{z}=[v_{z1}, \ldots, v_{zm}]^\intercal, z=1, \ldots, r, r<\min(p,m)$.\\

Note that the rank $r$ model is not immediately identifiable in terms of $\w_z$ and $\vv_z$ for different values of $z$. However, factorization techniques such as the singular value decomposition (SVD) can be used to obtain a unique representation of Equation (\ref{rankReq}).

\section{Estimation}
\label{s:inf}
Here we describe a general approach to estimating the coefficients for multi-way classification, in which an objective function is iteratively optimized over the weights in each dimension.  If the objective function at each iteration is convex, then the overall optimization problem is biconvex, and our approach can be framed as an Alternate Convex Search (ACS) algorithm \citep{gorski2007biconvex}. The objective functions of DWD, SVM, and several other linear classifiers are convex.  Therefore, we can iteratively optimize their objective functions using the ACS algorithm:
\[h(\y,\X,\B,\Theta) = h(\y,\vec{\X},\vec{\B},\Theta),\]
where $\vec{\X}:pm \times n$ and $\vec{\B}: pm \times 1$ correspond to the vectorized versions of $\X$ and $\B$, respectively.  The generic algorithm proceeds by iteratively estimating $\w$ and $\vv$ in $\B = \w \vv^\T$.

Below we give the algorithm in detail for multi-way DWD.  Details specific to the application of SVM to multi-way data are given in the Appendix~\ref{supp}.  In simulations and in practice, we find that both the basic version and multi-way version of DWD perform better than SVM, and so we focus on multi-way DWD hereafter. For the rank 1 multi-way model, the multi-way DWD algorithm is:

\textbf{Step 1:} \emph{Initialization}. Generate the random numbers $\tilde{w}_j^0, j=1,\ldots,p$ and $\tilde{v}_k^0,k=1,\ldots,m$ from a uniform distribution with range $0$ to $1$ and then set the initial values $\w^0=\frac{\tilde{\w}^0}{\parallel\tilde{\w}^0\parallel}$, and $\vv^0=\frac{\tilde{\vv}^0}{\parallel\tilde{\vv}^0\parallel}$ where $\tilde{\w}^0=(\tilde{w}_1^0,\ldots,\tilde{w}_p^0)^\intercal$ and $\tilde{\vv}^0=(\tilde{v}_1^0,\ldots,\tilde{v}_m^0)^\intercal$. Compute the median of the pairwise Euclidean distances between the two classes \citep{marron2007distance} and denote it as $D$.

\textbf{Step 2:} \emph{Iteration}. In the $(t+1)$th iteration step, first, using $\vv^t$, create a new dataset $\X^w$ where the observation for each subject $i$ is $\X_i^w=\X_{i}\cdot \vv^t$. Here $\X$ is the $n\times p\times m$ data array, so $\X_{i}$ is the $p\times m$ data matrix for subject $i$. Then update $\w^{t+1}$ by optimizing the DWD model to find the hyperplane defined by:
\begin{equation*}
f(\X_{i}^w)=w_{1}^{t+1}X_{i1}^w+w_{2}^{t+1}X_{i2}^w+\ldots+w_{p}^{t+1}X_{ip}^w.
\end{equation*}
Let $d_w$ denote the median of the pairwise Euclidean distances between the two classes in data $\X^w$; then the penalty parameter $C$ in the DWD model corresponding to $\X^w$ is set as $\frac{100*d_w^2}{D^2}$. Second, using $\w^{t+1}$ we apply a similar approach to update $\vv^{t+1}$.

\textbf{Step 3:} \emph{Convergence}. At the end of each iteration step, we compute the coefficients vector as $\vec{\B}^{t+1}=\vv^{t+1}\otimes \w^{t+1}$. If the Euclidean distance between $\vec{\B}^{t}$ and $\vec{\B}^{t+1}$ is less than a pre-specified threshold $\epsilon$, then the algorithm stops.
\\

For the rank $r$ model, we add an SVD procedure to assure the model is identifiable. The algorithm is:

\textbf{Step 1:} \emph{Initialization}. Generate the initial values for $w_{z,j}^0, j=1,\ldots,p$ and $v_{z,k}^0,k=1,\ldots,m$ and $z=1,\ldots,r$ from a uniform distribution with range $0$ to $1$. Compute the median of the pairwise Euclidean distances between the two classes \citep{marron2007distance} and denote it as $D$. Compute the coefficient matrix $\tilde{\B}^0=\w_{1}^0\cdot (\vv_{1}^{0})^{\intercal}+ \ldots +\w_{r}^0\cdot (\vv_{r}^{0})^{\intercal}$ where $\w_{z}^{0}=(w_{z,1}^0,\ldots,w_{z,p}^0)^\intercal$ and $\vv_{z}^{0}=(v_{z,1}^0,\ldots,v_{z,m}^0)^\intercal$, $z=1,\ldots,r$ and let $\B_{v}^0=\frac{\tilde{\B}^0}{\parallel\tilde{\B}^0\parallel}$. The subscript $v$ indicates that in the first iteration we will consider $\vv^0$ fixed to update $\w^1$; then we update $\vv^1$.

\textbf{Step 2:} \emph{Iteration}. In the $(t+1)$th iteration, compute the SVD of $\B_{v}^t$: $\UU_{p\times r}^t\Sigma_{r \times r}^t(\V_{m \times r}^{t})^{\intercal}$. Let ${\vv}_z^{t}$ be the $z$th column of $\V^t$. Create a new dataset $\X^w$ where the observation for each subject $i$ is $\X_i^w=\left((\X_{i}\cdot \vv_1^{t})^{\intercal},\ldots,(\X_{i}\cdot \vv_r^{t})^{\intercal}\right)^{\intercal}=\left(X_{i11}^w,\ldots,X_{i1p}^w,\ldots,X_{ir1}^w,\ldots,X_{irp}^w\right)^{\intercal}$ which is an $rp\times1$ vector. Then update $\tilde{\w}^{t+1}=\left((\tilde{\w}_1^{t+1})^\intercal,\ldots,(\tilde{\w}_r^{t+1})^\intercal\right)^{\intercal}$ where $\tilde{\w}_z^{t+1}=\left(\tilde{w}_{z1}^{t+1},\ldots,\tilde{w}_{zp}^{t+1}\right)^{\intercal}$, $z=1,\ldots,r$ by optimizing the DWD objective to find the hyperplane defined by:
\begin{equation*}
f(\X_{i}^w)=\tilde{w}_{11}^{t+1}X_{i11}^w+\tilde{w}_{12}^{t+1}X_{i12}^w+\ldots+\tilde{w}_{rp}^{t+1}X_{irp}^w.
\end{equation*}
Let $d_w$ denote the median of the pairwise Euclidean distances between the two classes in data $\X^w$; then the penalty parameter $C$ is set as $\frac{100 \cdot d_w^2}{D^2}$. Then update the coefficient matrix: \[\B_{w}^{t+1}=\tilde{w}_{1}^{t+1}\cdot (\vv_{1}^{t})^{\intercal}+ \ldots +\tilde{\w}_{r}^{t+1}\cdot (\vv_{r}^{t})^{\intercal}.\] 
After an SVD of $\B_{w}^{t+1}$, we use an analogous approach to update $\tilde{\vv}^{t+1}$, and obtain a new coefficient matrix \[\B_{v}^{t+1}=\w_{1}^{(t+1)}\cdot (\tilde{\vv}_{1}^{t+1})^{\intercal}+ \ldots +\w_{r}^{(t+1)}\cdot (\tilde{\vv}_{r}^{t+1})^{\intercal}.\]

\textbf{Step 3:} \emph{Convergence}. At the end of each iteration step, we compute the coefficients matrix $\B_{v}^{t+1}$. If the Euclidean distance between $\vec\B_{v}^{t}$ and $\vec\B_{v}^{t+1}$ is less than a pre-specified threshold $\epsilon$, then the algorithm stops.

\section{Simulation}
\label{s:Sim}

\subsection{Low rank model simulation and results}
We illustrate and compare the full DWD model to the proposed rank 1 multi-way model and rank $r$ multi-way model in a simulation study. The R package ``DWD" \citep{huang2012r} was used to fit the DWD model in each iteration step. Data were generated under several conditions, including different training sample sizes, different multi-way array dimensions, and different structural forms distinguishing the two classes.  
For all scenarios, a training dataset with sample size $n$ was generated, with two classes of equal size ($n_0=n_1=n/2$).  We consider the values $n=40$ and $n=100$. 
The predictors have the form of a two-way array of dimensions $p \times m$, and we generate data under three different array dimensions: $15\times 4$, $20\times 10$, and $500 \times 30$.

In each training dataset, $n_0$ samples corresponding to class $0$ were generated from a multivariate normal distribution $N(\mu_{0},\Sigma_{e0})$, where $\mu_{0}$ is a $pm\times 1$ vector and $\Sigma_{e0}=\sigma_{e0}^2 I_{pm\times pm}$. The other $n_1$ samples corresponding to class 1 were generated from a multivariate normal distribution $N(\mu_{1},\Sigma_{e1})$, where $\mu_{1}$ is a $pm\times 1$ vector and $\Sigma_{e1}=\sigma_{e1}^2 I_{pm\times pm}$.
Under this model the Bayes classification rule, which classifies a subject to class $i$ only if its multivariate density under class $i$ is highest, takes the form of a linear classifier in which the coefficients are proportional to the mean difference $\mu_1-\mu_0$.  This is easily shown by considering the difference in log density between the two classes. In practice the generative probability distribution is unknown, but this oracle rule may be used as a benchmark and motivates the three scenarios with different structure in the mean difference described below.

In the first structural form scenario, the data were generated from the full model. We set $\mu_0=(0, \ldots, 0)^\intercal$ and $\mu_1$ was generated from a multivariate normal distribution with mean zero and variance $\sigma_{s}^2 I_{pm\times pm}$. In the second scenario, the data were generated from the rank 1 model. We set $\mu_0=(0, \ldots, 0)^\intercal$. For $\mu_1$, we first generated $w$ from a multivariate normal distribution with mean zero and variance $\sigma_{w}^2 I_{p\times p}$ and $v$ from a multivariate normal distribution with mean zero and variance $\sigma_{v}^2 I_{m\times m}$. Then $\mu_1$ was determined by $v\otimes w$ where $\otimes$ denotes the Kronecker product. In the third scenario, the data were generated from the rank 2 model. We first generated $w_0$ from a multivariate normal distribution with mean zero and variance $\sigma_{w_0}^2 I_{p\times p}$ and $v_0$ from a multivariate normal distribution with mean zero and variance $\sigma_{v_0}^2 I_{m\times m}$ and then $\mu_0$ was determined by $v_0\otimes w_0$. Similarly, $\mu_1$ was determined by $v_1\otimes w_1$ where $w_1$ was generated from a multivariate normal distribution with mean zero and variance $\sigma_{w_1}^2 I_{p\times p}$ and $v_1$ was from a multivariate normal distribution with mean zero and variance $\sigma_{v_1}^2 I_{m\times m}$.

Under each scenario, corresponding test data were generated from the same distributions as the training data with sample sizes $n_0=n_1=50$.  For each method we assess its misclassification rate on the test data, and the correlation of the estimated hyperplane and the true (Bayes) hyperplane.  The signal-to-noise ratio was adjusted for each scenario such that the misclassification rate of the full model is around $20\%$ when $n=40$. Each scenario was replicated $100$ times.

The results in Table \ref{ta:sim1} and Figure \ref{simulation} show that the model with the best performance (in terms of the misclassification rates and correlation with the truth) is the model from which the data were generated under each scenario. When the sample size increases from 40 to 100, the misclassification rates are lower and the correlations between the estimated and the true hyperplanes are higher. From Figure \ref{simulation}, we also observe that when the dimensions increase, the differences in misclassification rate between the appropriate model and alternative models increase.

\begin{table}[!h]
	\tblcaption{Simulation results: ``Mis" is the misclassification rate and ``SE(Mis)" is the standard error of the misclassification rate across the 200 simulated datasets. ``Cor" is the correlation between the estimated linear hyperplane and the true hyperplane. ``SE(Cor)" is the standard error of the correlation.  The best correlation and misclassification rate in each row are given in \textbf{bold} font.}
	\centering
	\resizebox{\textwidth}{!}{
		{\tabcolsep=4.25pt
		\begin{tabular}{cccccccccccccccccc}
			\tblhead{&&&&\multicolumn{4}{c}{full model}&&\multicolumn{4}{c}{rank 1 model}&&\multicolumn{4}{c}{rank 2 model}\\
			\cline{5-8}\cline{10-13}\cline{15-18}
			$n$&\text{dimension}&\text{true model}&&Mis&SE(Mis)&Cor&SE(Cor)&&Mis&SE(Mis)&Cor&SE(Cor)&&Mis&SE(Mis)&Cor&SE(Cor)}\\
		    40  & $15\times 4$ & full &  & \textbf{0.202} & 0.004 & \textbf{0.672} & 0.005 &  & 0.288 & 0.005 & 0.452 & 0.007 &  & 0.238 & 0.004 & 0.575 & 0.006 \\ 
		      &  & rank 1 &  & 0.196 & 0.010 & 0.669 & 0.012 &  & \textbf{0.156} & 0.009 & \textbf{0.798} & 0.014 &  & 0.178 & 0.009 & 0.720 & 0.012 \\ 
		      &  & rank 2 &  & 0.207 & 0.008 & 0.664 & 0.009 &  & 0.195 & 0.008 & 0.700 & 0.012 &  & \textbf{0.194} & 0.008 & \textbf{0.710} & 0.010 \\ 
		      & $20\times 10$ & full &  & \textbf{0.205} & 0.003 & \textbf{0.545} & 0.004 &  & 0.341 & 0.004 & 0.272 & 0.004 &  & 0.296 & 0.004 & 0.358 & 0.004 \\ 
		      &  & rank 1 &  & 0.212 & 0.008 & 0.530 & 0.008 &  & \textbf{0.127} & 0.007 & \textbf{0.799} & 0.010 &  & 0.159 & 0.008 & 0.689 & 0.009 \\ 
		      &  & rank 2 &  & 0.209 & 0.006 & 0.535 & 0.006 &  & 0.179 & 0.007 & 0.635 & 0.011 &  & \textbf{0.166} & 0.006 & \textbf{0.664} & 0.008 \\ 
		      & $500\times 30$ & full &  & \textbf{0.202} & 0.003 & \textbf{0.206} & 0.001 &  & 0.429 & 0.004 & 0.046 & 0.001 &  & 0.400 & 0.004 & 0.064 & 0.001 \\ 
		      &  & rank 1 &  & 0.215 & 0.005 & 0.200 & 0.002 &  & \textbf{0.008} & 0.001 & \textbf{0.692} & 0.005 &  & 0.026 & 0.003 & 0.545 & 0.005 \\ 
		      &  & rank 2 &  & 0.212 & 0.004 & 0.201 & 0.001 &  & 0.049 & 0.003 & 0.445 & 0.005 &  & \textbf{0.040} & 0.003 & \textbf{0.493} & 0.006 \\ 
		    100 & $15\times 4$ & full &  & \textbf{0.154} & 0.046 & \textbf{0.821} & 0.044 &  & 0.247 & 0.055 & 0.553 & 0.078 &  & 0.194 & 0.051 & 0.702 & 0.059 \\ 
		      &  & rank 1 &  & 0.159 & 0.122 & 0.801 & 0.127 &  & \textbf{0.138} & 0.123 & \textbf{0.900} & 0.150 &  & 0.150 & 0.123 & 0.842 & 0.138 \\ 
		      &  & rank 2 &  & 0.165 & 0.098 & 0.804 & 0.094 &  & 0.160 & 0.100 & 0.810 & 0.118 &  & \textbf{0.152} & 0.101 & \textbf{0.846} & 0.107 \\ 
		      & $20\times 10$ & full &  & \textbf{0.137} & 0.040 & \textbf{0.720} & 0.033 &  & 0.292 & 0.057 & 0.360 & 0.050 &  & 0.233 & 0.051 & 0.477 & 0.046 \\ 
		      &  & rank 1 &  & 0.148 & 0.094 & 0.704 & 0.099 &  & \textbf{0.086} & 0.068 & \textbf{0.921} & 0.063 &  & 0.108 & 0.082 & 0.842 & 0.082 \\ 
		      &  & rank 2 &  & 0.146 & 0.081 & 0.709 & 0.074 &  & 0.135 & 0.073 & 0.762 & 0.092 &  & \textbf{0.106} & 0.068 & \textbf{0.853} & 0.077 \\ 
		      & $500\times 30$ & full &  & \textbf{0.096} & 0.030 & \textbf{0.317} & 0.007 &  & 0.385 & 0.050 & 0.072 & 0.007 &  & 0.341 & 0.050 & 0.100 & 0.008 \\ 
		      &  & rank 1 &  & 0.114 & 0.060 & 0.309 & 0.038 &  & \textbf{0.001} & 0.004 & \textbf{0.853} & 0.037 &  & 0.005 & 0.011 & 0.734 & 0.050 \\ 
		      &  & rank 2 &  & 0.107 & 0.051 & 0.311 & 0.029 &  & 0.010 & 0.014 & 0.618 & 0.054 &  & \textbf{0.002} & 0.006 & \textbf{0.749} & 0.042 \\ 
			\lastline
		\end{tabular}}
	}
	\label{ta:sim1}
\end{table}

\subsection{High rank model simulation and results}

\begin{figure}[!h]
	\begin{center}
		\includegraphics[width = \textwidth]{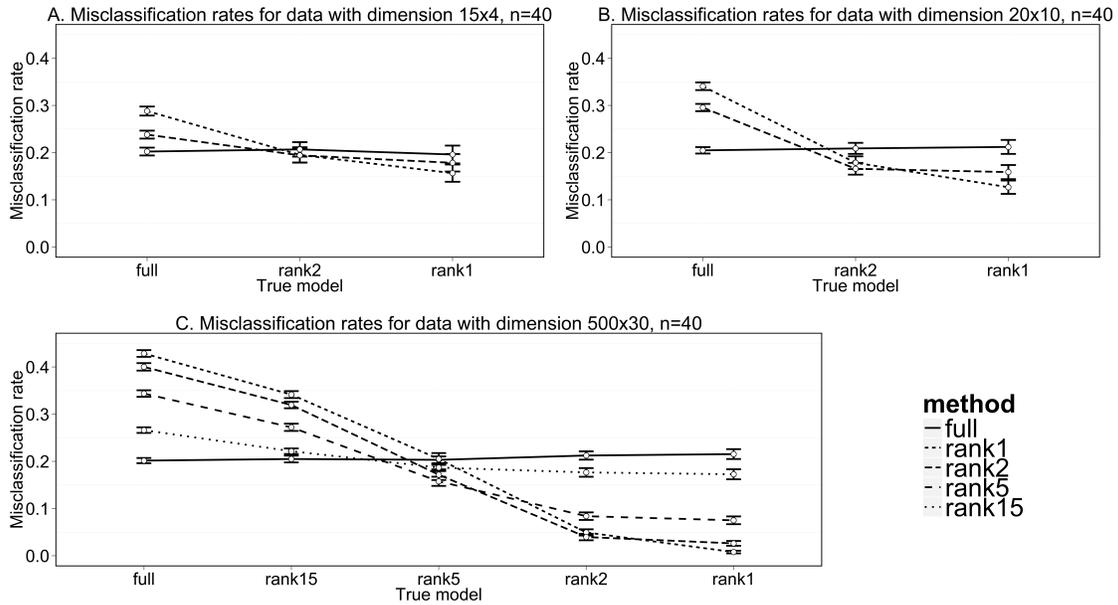}
	\end{center}
	\caption{Misclassification rates with bars for $\pm$ 1.96 standard errors (across the 200 simulated datasets) under each simulation scenario.}
\label{simulation}
\end{figure}

When the dimension of the multi-way structure is $500\times 30$, the effective dimensions of the full model, rank 1 model and rank 2 model are $500\times 30=15000$, $500+30=530$ and $500+499+30+29=1058$ respectively. By this measure, the difference between the full model and the rank 2 model is quite large. In order to evaluate the performance of rank $r$ models with effective dimension in between, additional simulations were done for the multi-way structure with dimension $500\times 30$. We added two more scenarios where the simulated datasets were generated under the rank 5 and rank 15 models and then applied the rank 5 and rank 15 models to all datasets generated under the $500\times 30$ structure. The results are shown in Figure \ref{simulation}, and illustrate how different low-rank approximations serve as a flexible compromise between the rank 1 and full models.  In particular, as $r$ increases, the performance of the rank $r$ model approaches that of the full model.  

\section{Real data analysis}
The proposed methods were illustrated in two real data examples.

\subsection{Magnetic Resonance Spectroscopy (MRS) data} \label{MRS_sec}
We consider Magnetic Resonance Spectroscopy (MRS) data for a clinical research project that enrolled patients with Spinocerebellar Ataxia Type 1 (SCA1) and healthy controls of similar age and sex distribution. MRS is a non-invasive method using magnetic resonance imaging to quantify neurochemicals. Here it was used to examine differences between patients and controls, and ultimately to track changes in the brains of patients as the disease progresses. Participants were imaged in a 3 Tesla scanner and the neurochemicals that were quantified included ascorbate (Asc), $\gamma$-aminobutyric acid (GABA), glucose (Glc), glutamate (Glu), glutathione (GSH), myo-inositol (Ins), scyllo-inositol (sIns), N-acetyl-aspertate (NAA), total choline (Pcho+GPC), total creatine (Cr+PCr), total NAA (NAA+NAAG), glutamate plus glutamine (Glu+Gln), and glucose plus taurine (Glc+Tau).

\begin{figure}[!h]
	\begin{center}
		\includegraphics[width = \textwidth]{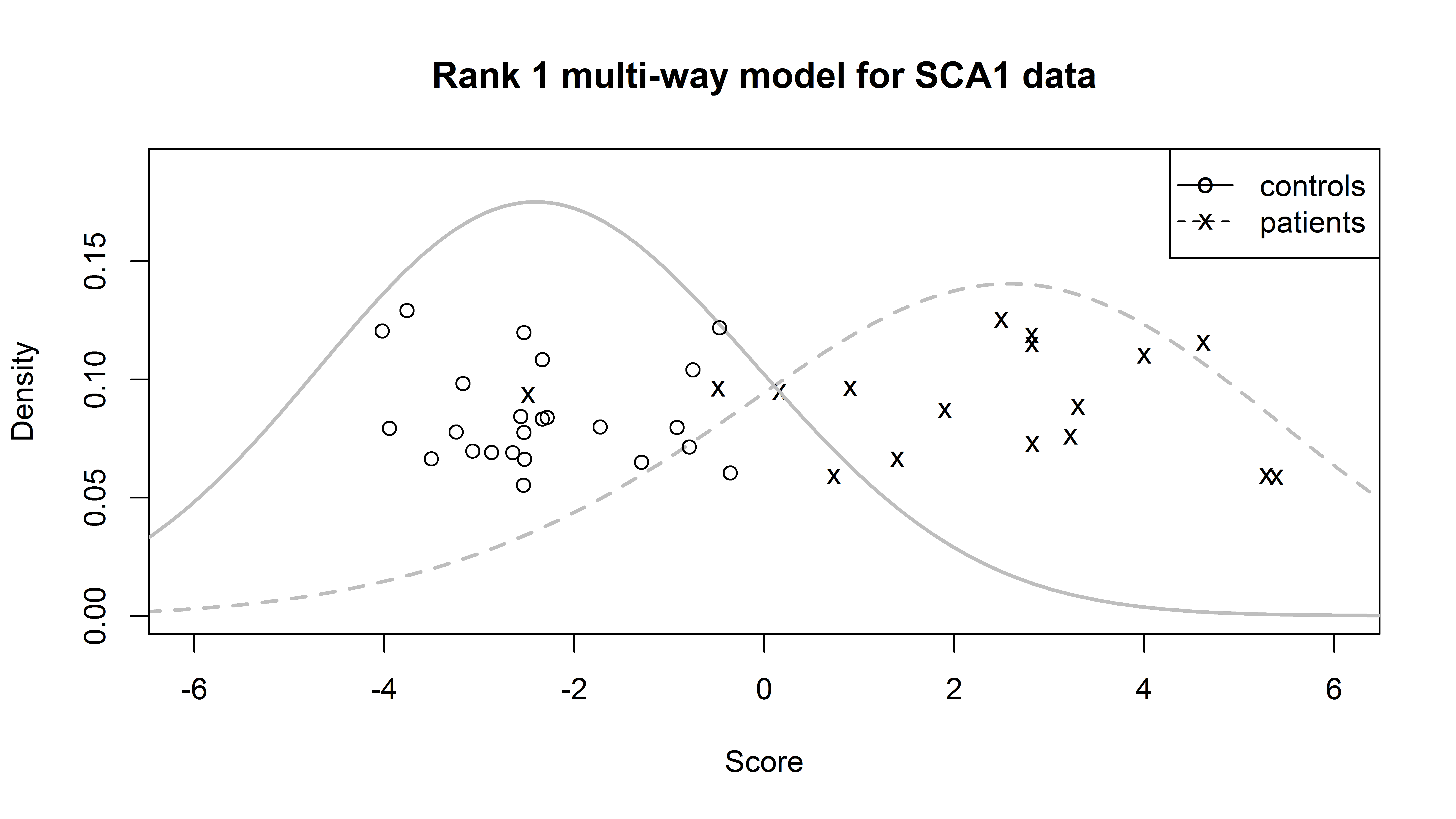}
	\end{center}
	\caption{Rank 1 multi-way DWD scores under leave-one-out cross-validation for controls and patients, with a kernel density estimate for each group.}
\label{MRSscore}
\end{figure}

\begin{figure}[!h]
	\begin{center}
		\includegraphics[width = \textwidth]{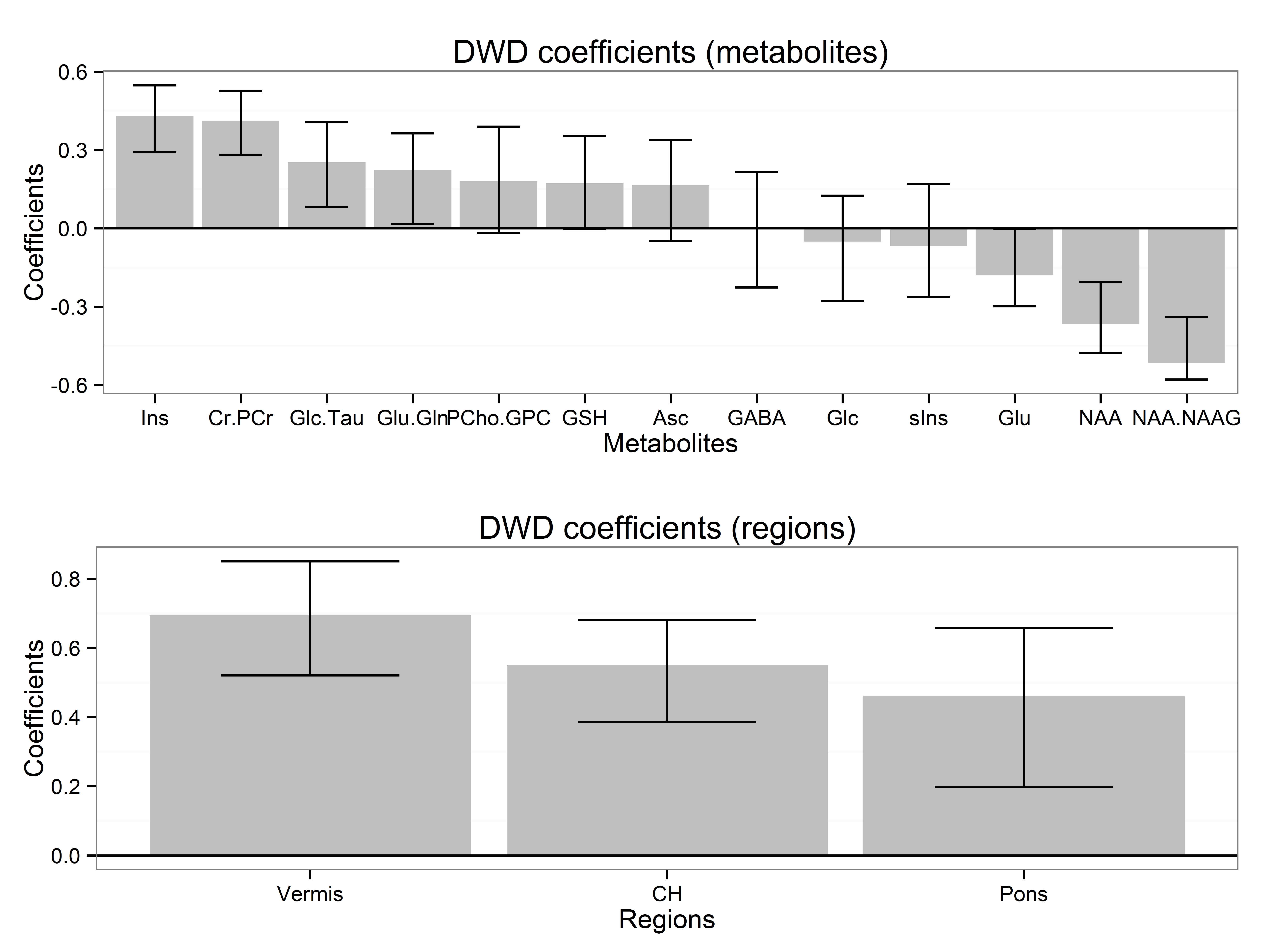}
	\end{center}
	\caption{Rank 1 multi-way DWD weights for metabolites (top panel) and regions (bottom panel), with 95\% confidence intervals generated from $5000$ bootstrap samples.}
\label{MRSCoeffs}
\end{figure}

There were 17 patients and 24 controls enrolled in this study. The concentrations of the same set of metabolites were measured in three different brain regions (Pons, Cerebral Hemisphere, and Vermis), yielding data with three dimensions: \emph{participants} $\times$ \emph{metabolites} $\times$ \emph{regions}. Thus, the data have a multi-way structure. We compared misclassification rates for the full model and the rank 1 multi-way DWD model by leave-one-out cross validation, which is robust to over-fitting. Each sample was separately left out of the estimation (to be the validation set), all the other samples were used as training samples to construct the model, and then the model was tested on the left-out sample. The two models gave the same leave-one-out misclassification rate of $4.88\%$ and similar t-statistics (8.815 vs. 8.354 for rank 1 and full, respectively). The t-statistic corresponded to testing the null hypothesis that the mean DWD scores of the two groups are the same, where the DWD score for each sample is calculated from Equation \ref{MultiWayPlane}. The DWD scores under the rank 1 multi-way DWD model are shown in Figure \ref{MRSscore}, which shows that the patients and controls are well separated. Note that one misclassified case patient scored in the middle of the controls; this was a presymptomatic patient diagnosed by genetic screening (rather than presentation of clinical symptoms), so it is reasonable that the rank 1 multiway DWD model could not classify this patient correctly. The rank 1 multi-way model estimated a single weight for each metabolite ($\vv$) and a single weight for each region ($\w$), thus it has a simpler interpretation compared to the full model. In order to estimate the 95\% confidence intervals of the estimated weights, 5000 bootstrap samples were generated. For each bootstrap sample, 17 patients and 24 controls were resampled with replacement from the original 17 patients and 24 controls separately. Then the model was fit to the bootstrap sample to get the estimated weights. The 95\% confidence interval was constructed based on the 2.5\% and 97.5\% quantiles of all the estimated weights based on the bootstrap samples. The estimated weights and their 95\% bootstrap confidence intervals are shown in Figure \ref{MRSCoeffs}. The metabolites with large absolute weights are considered important in distinguishing the ataxia patients from the healthy controls.

\subsection{Gene Time Course Data}

We applied multi-way DWD to classify clinical response to treatment for Multiple Sclerosis (MS) patients based on gene expression time course data.  These data were originally described in \citet{baranzini2005}.  Fifty-three patients were given recombinant human interferon beta (rIFN$\beta$), which is often used to control the symptoms of MS.  Gene expression was measured for 76 genes of interest before treatment (baseline) and at 6 follow-up time points over the next two years (3 months, 6 months, 9 months, 12 months, 18 months, 24 months), yielding a 3-way data array: \emph{patients} $\times$ \emph{genes} $\times$ \emph{times}.  Afterward, patients were classified into good responders or poor responders to rIFN$\beta$ based on clinical characteristics.  Efficient classification of good and poor responders from the gene expression data is desired, for example to guide treatment decisions and to better understand the IFN$\beta$ mechanism.  The raw data are publicly available as a supplemental file to \citet{baranzini2005}.

We consider rank-$r$ multi-way DWD to classify good and poor responders for each of $r=1,\hdots,7$.  The seven models were compared via leave-one-out cross validation estimation of the mis-classification rate. The rank-$1$ model, with a single weight for each gene and for each time point, outperformed the others with the highest t-statistic ($t=7.58$) and lowest misclassification rate ($16.9\%$) under cross validation.  The full model, with a distinct coefficient for each gene $\times$ time pair, had a t-statistic of $5.38$ and a misclassification rate of $22.6\%$ under cross-validation.  

\begin{figure}[!h]
	\begin{center}
		\includegraphics[width = \textwidth]{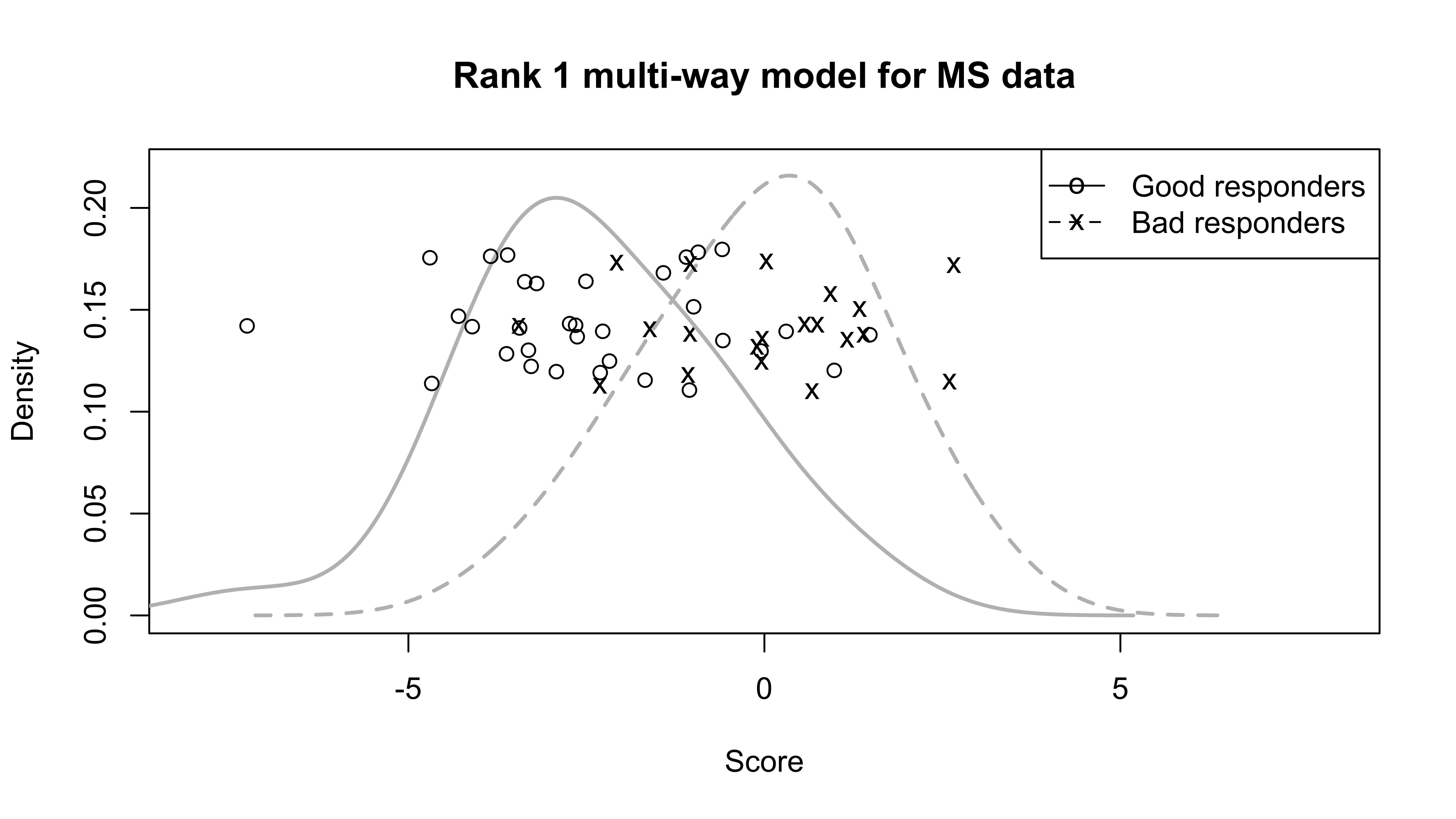}
	\end{center}
	\caption{Rank 1 multi-way DWD scores under leave-one-out cross-validation for good and poor treatment responders, with a kernel density estimate for each group.}
\label{GeneScores}
\end{figure}

\begin{figure}[!h]
	\begin{center}
		\includegraphics[width = \textwidth]{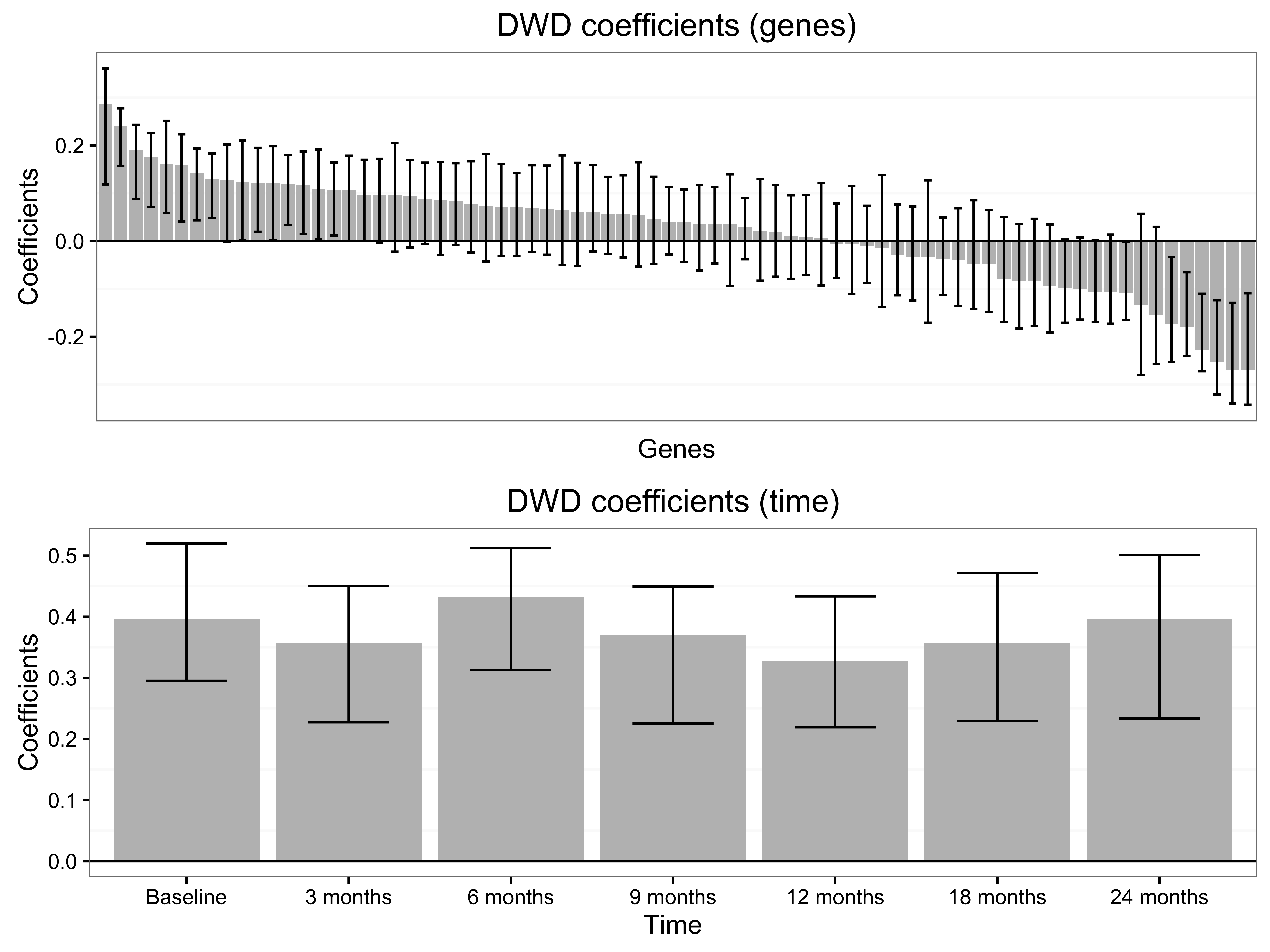}
	\end{center}
	\caption{Rank 1 multi-way DWD weights for genes (top panel) and time points (bottom panel), with 95\% confidence intervals generated from $5000$ bootstrap samples.}
\label{GeneCoeffs}
\end{figure}

The DWD scores under leave-one-out cross validation for the rank-1 multi-way model are shown in Figure~\ref{GeneScores}. This shows substantial but not perfect discrimination between the good and poor responder groups.  The coefficient estimates for each gene and each time point, with 95\% bootstrap confidence intervals, are shown in Figure \ref{GeneCoeffs}.  The four genes with the highest coefficient were Jak1, Caspase.9, STAT3, and   IFN.gRa; the four genes with the highest negative coefficient were FAS, NFkBIA, IRF6, and ITGB2. The coefficients across time had little variability and no noticeable patterns.  This suggests that the distinction between good and poor responders is not driven by changes to gene expression in response to INF$\beta$, but rather by baseline differences in expression that can be quantified more precisely over multiple time points.  This agrees with the results in \citet{baranzini2005}, who conducted an analysis of variance (ANOVA) for each gene and report several significant response and time effects but no response*time interactions.  

An alternative approach to classifying subjects based on gene expression time-course data is described in \citet{zhang2013}.  Their method identifies an optimal direction in \emph{time} for each gene using Fisher's LDA, and then applies SVM or another high-dimensional classifier to the projections of each gene on its optimal direction.  This approach is appropriate when discriminative patterns over time are different for each gene, but it does not explicitly capture patterns that are shared across multiple genes.  They assess classification accuracy using the same IFN$\beta$ dataset described above and achieve a minimum cross-validation error rate of approximately $26\%$, greater than the error rate of $16.9\%$ achieved by multi-way DWD.

\section{Discussion}
\label{s:discuss}
Although data with multi-way structure is common in biomedical research, little work has addressed classification of categorical outcomes from high-dimensional multi-way data. In this article, we have proposed a general framework to extend linear classification methods to multi-way data. We mainly focus on the multi-way DWD model because of its ability to overcome the data-piling problem of SVM and its good performance in simulations. Both the simulation and real data analysis results show that the multi-way model can improve classification accuracy when the underlying true model has a multi-way structure and can provide a simple and straightforward interpretation.

While the simple rank 1 classification model performs well in our applications, it may perform poorly for others.  The simulations in Section~\ref{s:Sim} clearly demonstrate that low-rank models can be a poor approximation if the underlying signal distinguishing the two classes does not have multi-way structure.  Therefore, we advise that the rank of the model should not be decided upon blindly; rather, multiple models should be considered and compared via cross-validated errors or other means.

The methodology described in Sections~\ref{sec:meth} and ~\ref{s:inf} may be extended in several ways.  While we implement multi-way classification for a binary outcome, the framework can also be used to extend multi-category classification methods such as multiclass DWD \citep{huang2013}.  Also in our implementation we focus on three-way data: \emph{samples} $\times$ \emph{dim}$_1$ $\times$ \emph{dim}$_2$.  The general framework and iterative estimation technique may be extended to more dimensions; for higher-order arrays the coefficients may be represented as a rank-$r$ PARAFAC decomposition.  Sparse versions of SVM and DWD have been developed, in which negligible coefficients are shrunk to $0$ \citep{bi2003,wang2015}.  Sparse multi-way classification, in which some of the weights in each dimension are shrunk to $0$, is another direction of future development.

\section*{Acknowledgments}

We thank G$\ddot{\text{u}}$lin $\ddot{\text{O}}$z, Uzay Emir, and Dinesh Deelchand for providing the data and feedback for the MRS application described in Section~\ref{MRS_sec}. We also thank Hanwen Huang for his help and advice regarding the DWD package for R.\\
{\it Conflict of Interest}: None declared.

\section*{Funding}

This work was supported by the National Institutes of Health grant ULI RR033183/KL2
RR0333182 [to EFL] and grant 1R01NS080816-01A1 [supporting TL and LEE].

\appendix

\section{Multi-Way SVM}
\label{supp}
Details specific to the application of SVM to multi-way data are given below.
For the rank 1 multi-way model, the multi-way SVM algorithm is:

\textbf{Step 1:} \emph{Initialization.} generate the random numbers $w_j^0, j=1,\ldots,p$ and $v_k^0,k=1,\ldots,m$ from a uniform distribution with range $0$ to $1$ and then set the initial values $\w^0=(w_1^0,\ldots,w_p^0)^\intercal$ and $\vv^0=(v_1^0,\ldots,v_m^0)^\intercal$.

\textbf{Step 2:} \emph{Iteration.}  in the $(t+1)$th iteration step, first, standardize $\vv^t$ by $\vv^t=\frac{\vv^t}{\|\vv^t\|}$ and then fix $\vv^t$, and create a new dataset $\X^w$ where the observation for each subject $i$ is $\X_i^w=\X_{i}\cdot \vv^t$. Here $\X$ is the $n\times p\times m$ data array, so $\X_{i}$ is the $p\times m$ data matrix for subject $i$. Then update $\w^{t+1}$ by optimizing the SVM model to find the hyperplane that:
\begin{equation*}
f(\X_{i}^w)=w_{1}^{t+1}X_{i1}^w+w_{2}^{t+1}X_{i2}^w+\ldots+w_{p}^{t+1}X_{ip}^w.
\end{equation*}
Second, we standardize and fix $\w^{t+1}$ and then apply a similar approach to update $\vv^{t+1}$.

\textbf{Step 3:} \emph{Convergence.} in the end of each iteration step, we compute the coefficients vector by $\bc^{t+1}=\vv^{t+1}\otimes \w^{t+1}$. If the Euclidean difference between $\bc^{t}$ and $\bc^{t+1}$ is less than a pre-specified threshold $\epsilon$, then the algorithm stops.

For the rank $r$ model, we add an SVD procedure to assure the model is identifiable. The algorithm is:

\textbf{Step 1:} Initialization: generate the initial values for $w_{z,j}^0, j=1,\ldots,p$ and $v_{z,k}^0,k=1,\ldots,m$ and $z=1,\ldots,r$ from a uniform distribution with range $0$ to $1$. Compute the coefficient matrix $\tilde{\B}^0=\w_{1}^0\cdot (\vv_{1}^{0})^{\intercal}+ \ldots +\w_{r}^0\cdot (\vv_{r}^{0})^{\intercal}$ where $\w_{z}^{0}=(w_{z,1}^0,\ldots,w_{z,p}^0)^\intercal$ and $\vv_{z}^{0}=(v_{z,1}^0,\ldots,v_{z,m}^0)^\intercal$, $z=1,\ldots,r$.

\textbf{Step 2:} Iteration: In the $(t+1)$th iteration, compute the SVD of $\B_{v}^t$: $\UU_{p\times r}^t\Sigma_{r \times r}^t(\V_{m \times r}^{t})^{\intercal}$. Let ${\vv}_z^{t}$ be the $z$th column of $\V^t$. Create a new dataset $\X^w$ where the observation for each subject $i$ is $\X_i^w=\left((\X_{i}\cdot \vv_1^{t})^{\intercal},\ldots,(\X_{i}\cdot \vv_r^{t})^{\intercal}\right)^{\intercal}=\left(X_{i11}^w,\ldots,X_{i1p}^w,\ldots,X_{ir1}^w,\ldots,X_{irp}^w\right)^{\intercal}$ which is an $rp\times1$ vector. Then update $\tilde{\w}^{t+1}=\left((\tilde{\w}_1^{t+1})^\intercal,\ldots,(\tilde{\w}_r^{t+1})^\intercal\right)^{\intercal}$ where $\tilde{\w}_z^{t+1}=\left(\tilde{w}_{z1}^{t+1},\ldots,\tilde{w}_{zp}^{t+1}\right)^{\intercal}$, $z=1,\ldots,r$ by optimizing the SVM objective to find the hyperplane defined by:
\begin{equation*}
f(\X_{i}^w)=\tilde{w}_{11}^{t+1}X_{i11}^w+\tilde{w}_{12}^{t+1}X_{i12}^w+\ldots+\tilde{w}_{rp}^{t+1}X_{irp}^w.
\end{equation*}
 Then update the coefficient matrix: \[\B_{w}^{t+1}=\tilde{w}_{1}^{t+1}\cdot (\vv_{1}^{t})^{\intercal}+ \ldots +\tilde{\w}_{r}^{t+1}\cdot (\vv_{r}^{t})^{\intercal}.\] 
After an SVD of $\B_{w}^{t+1}$, we use an analogous approach to update $\tilde{\vv}^{t+1}$, and obtain a new coefficient matrix \[\B_{v}^{t+1}=\w_{1}^{(t+1)}\cdot (\tilde{\vv}_{1}^{t+1})^{\intercal}+ \ldots +\w_{r}^{(t+1)}\cdot (\tilde{\vv}_{r}^{t+1})^{\intercal}.\]

\textbf{Step 3:} Convergence: in the end of each iteration step, we compute the coefficients matrix $\B_{v}^{t+1}$. If the Euclidean difference between $\vec\B_{v}^{t}$ and $\vec\B_{v}^{t+1}$ is less than a pre-specified threshold $\epsilon$, then the algorithm stops.

Note that the convergence of the multi-way SVM model depends more on the starting random seed than the multi-way DWD model. So we try multiple starting random seeds and select the solution with the largest objective function value as the best solution.

The comparison of DWD and SVM was shown in Table \ref{ta:simsvm}. Only the results based on the model from which the data were generated were reported. For example, when the data were generated from the rank 1 model, then the comparison was between the rank 1 multi-way DWD model and the rank 1 multi-way SVM model.

\begin{table}[!h]
	\tblcaption{Simulation results: ``Mis" is the misclassification rate and ``SE(Mis)" is the standard error of the misclassification rate across the 200 simulated datasets. ``Cor" is the absolute value of the correlation between the estimated linear hyperplane and the true hyperplane (we take the absolute value because SVM does not define a positive and negative class). ``SE(Cor)" is the standard error of the absolute correlation. }
	\centering
	\resizebox{\textwidth}{!}{
		{\tabcolsep=4.25pt
			\begin{tabular}{cccccccccccc}
				\tblhead{&&&\multicolumn{4}{c}{DWD}&&\multicolumn{4}{c}{SVM}\\
					\cline{4-7}\cline{9-12}
					\text{dimension}&\text{true model}&&Mis&SE(Mis)&Cor&SE(Cor)&&Mis&SE(Mis)&Cor&SE(Cor)}\\
				  $15\times 4$& full &  & 0.202 & 0.004 & 0.672 & 0.005 &  & 0.239 & 0.004 & 0.575 & 0.006 \\ 
				  & rank 1 &  & 0.156 & 0.009 & 0.799 & 0.014 &  & 0.211 & 0.010 & 0.620 & 0.015 \\ 
				  & rank 2 &  & 0.194 & 0.008 & 0.710 & 0.010 &  & 0.244 & 0.008 & 0.550 & 0.011 \\ 
				  $20\times 10$& full &  & 0.205 & 0.003 & 0.545 & 0.004 &  & 0.218 & 0.004 & 0.511 & 0.004 \\ 
				  & rank 1 &  & 0.127 & 0.007 & 0.799 & 0.010 &  & 0.206 & 0.009 & 0.556 & 0.014 \\ 
				  & rank 2 &  & 0.166 & 0.006 & 0.664 & 0.008 &  & 0.228 & 0.007 & 0.490 & 0.009 \\ 
				  $500\times 30$& full &  & 0.202 & 0.003 & 0.206 & 0.001 &  & 0.201 & 0.003 & 0.206 & 0.001 \\ 
				  & rank 1 &  & 0.008 & 0.001 & 0.692 & 0.005 &  & 0.024 & 0.003 & 0.577 & 0.007 \\ 
				  & rank 2 &  & 0.040 & 0.003 & 0.493 & 0.006 &  & 0.066 & 0.005 & 0.415 & 0.006 \\ 
				\lastline
			\end{tabular}}
		}
		\label{ta:simsvm}
	\end{table}

\bibliographystyle{biorefs}
\bibliography{multiway}

\end{document}